\documentclass{PoS}

\usepackage[utf8]{inputenc}
\usepackage{subfigure}
\usepackage{booktabs}
\usepackage{multirow}
\usepackage[permil]{overpic}
\usepackage[load-configurations=abbreviations]{siunitx}
\usepackage{upgreek}
\title{Upgrade of the ALICE Inner Tracking System}

\ShortTitle{Upgrade of the ALICE Inner Tracking System}

\author{\speaker{Felix Reidt for the ALICE collaboration}%
        \thanks{My work is supported by the Wolfgang-Gentner programme of the Bundesministerium für Bildung und Forschung (BMBF).}\\
       CERN, 1210 Geneva 23, Switzerland and \\
       Physikalisches Institut, Ruprecht-Karls-Universitaet Heidelberg,\\
       Im Neuenheimer Feld 226, 69120 Heidelberg, Germany\\
       E-mail: \email{felix.reidt@cern.ch}}

\abstract{During the Long Shutdown 2 of the LHC in 2018/2019, the ALICE experiment plans the installation of a novel Inner Tracking System. It will replace the current six layer detector system with a seven layer detector using Monolithic Active Pixel Sensors. The upgraded Inner Tracking System will have significantly improved tracking and vertexing capabilities, as well as readout rate to cope with the expected increased Pb-Pb luminosity of the LHC. The choice of Monolithic Active Pixel Sensors has been driven by the specific requirements of ALICE as a heavy ion experiment dealing with rare processes at low transverse momenta. This leads to stringent requirements on the material budget of \SI{0.3}{\%~X/X_{0}} per layer for the three innermost layers. Furthermore, the detector will see large hit densities of $\sim\SI{19}{cm^{-2}/event}$ on average for minimum-bias events in the inner most layer and has to stand a moderate maximum total ionising dose of \SI{700}{krad} and a non-ionising energy loss of $\SI{1e13}{\SI{1}{MeV~n_{eq}/cm^{2}}}$. The Monolithic Active Pixel Sensor detectors are manufactured using the TowerJazz $\SI{0.18}{\upmu{}m}$ CMOS Imaging Sensor process on wafers with a high-resistivity epitaxial layer.

This contribution summarises the recent R\&D activities and focuses on results on the large-scale pixel sensor prototypes.
}

\FullConference{The 23rd International Workshop on Vertex Detectors\\
		 15-19 September 2014\\
		 Macha Lake, The Czech Republic}

\begin{document}
\section{Introduction}
ALICE (A Large Ion Collider Experiment) \cite{Aamodt2008} is a general-purpose, heavy-ion experiment at the CERN LHC. Its main goal is to study the physics properties of the Quark-Gluon Plasma. During the Long Shutdown 2 (LS2) of the LHC in 2018/2019, ALICE will undergo a major upgrade in order to significantly enhance its physics capabilities, in particular for high precision measurements of rare processes at low transverse momenta $p_\text{T}$.
\subsection{ALICE Upgrade}
The ALICE upgrade programme~\cite{Alice2014a} during LS2 is based on a combination of detector upgrades improving their physics performance and preparing them for a significant luminosity increase to $L=\SI{6e27}{\cm^{-2}\s^{-1}}$ for nucleus-nucleus (A-A) collisions. The increased luminosity will lead to a Pb-Pb interaction rate of about \SI{50}{kHz}. The study of rare probes at low $p_\text{T}$ in heavy-ion collisions makes triggering inefficient due to the large combinatorial background~\cite{Alice2012}. Thus, the upgraded experimental apparatus is designed to readout all Pb-Pb interactions, accumulating events corresponding to an integrated luminosity of more than \SI{10}{nb^{-1}}. This minimum-bias data sample will provide an increase in statistics by about a factor 100 with respect to the programme until LS2. The upgraded detector will provide improved vertexing and tracking capabilities at low $p_\text{T}$. In summary, the detector upgrade consists of the following sub-system upgrades:
\begin{itemize}
\item Reduction of the beam-pipe radius from \SI{29.8}{mm} to \SI{19.8}{mm} allowing the inner layer of the central barrel silicon tracker to be moved closer to the interaction point.
\item New high-resolution, high-granularity, low material budget silicon trackers:
  \begin{itemize}
  \item Inner Tracking System (ITS)~\cite{Alice2014b} covering mid-pseudo-rapidity ($-1.2<\eta<1.2$).
  \item Muon Forward Tracker (MFT)~\cite{Alice2013a} covering forward pseudo-rapidity ($-3.6<\eta<2.45$).
  \end{itemize}
\item The wire chambers of the Time Projection Chamber (TPC) will be replaced by GEM detectors and new electronics will be installed in order to allow for a continuous readout~\cite{Alice2013b}.
\item Upgrade of the forward trigger detectors and the Zero Degree Calorimeter~\cite{Alice2013c}.
\item Upgrade of the readout electronics of the Transition Radiation Detector (TRD), Time-Of-Flight (TOF) detector, PHOton Spectrometer(PHOS) and Muon Spectrometer for high rate operation~\cite{Alice2013c}.
\item Upgrade of online and offline systems (O$^2$ project)~\cite{Alice2014a} in order to cope with the expected data volume.
\end{itemize}

\section{ALICE ITS Upgrade}
\begin{figure}
  \begin{overpic}[width=0.51\textwidth]{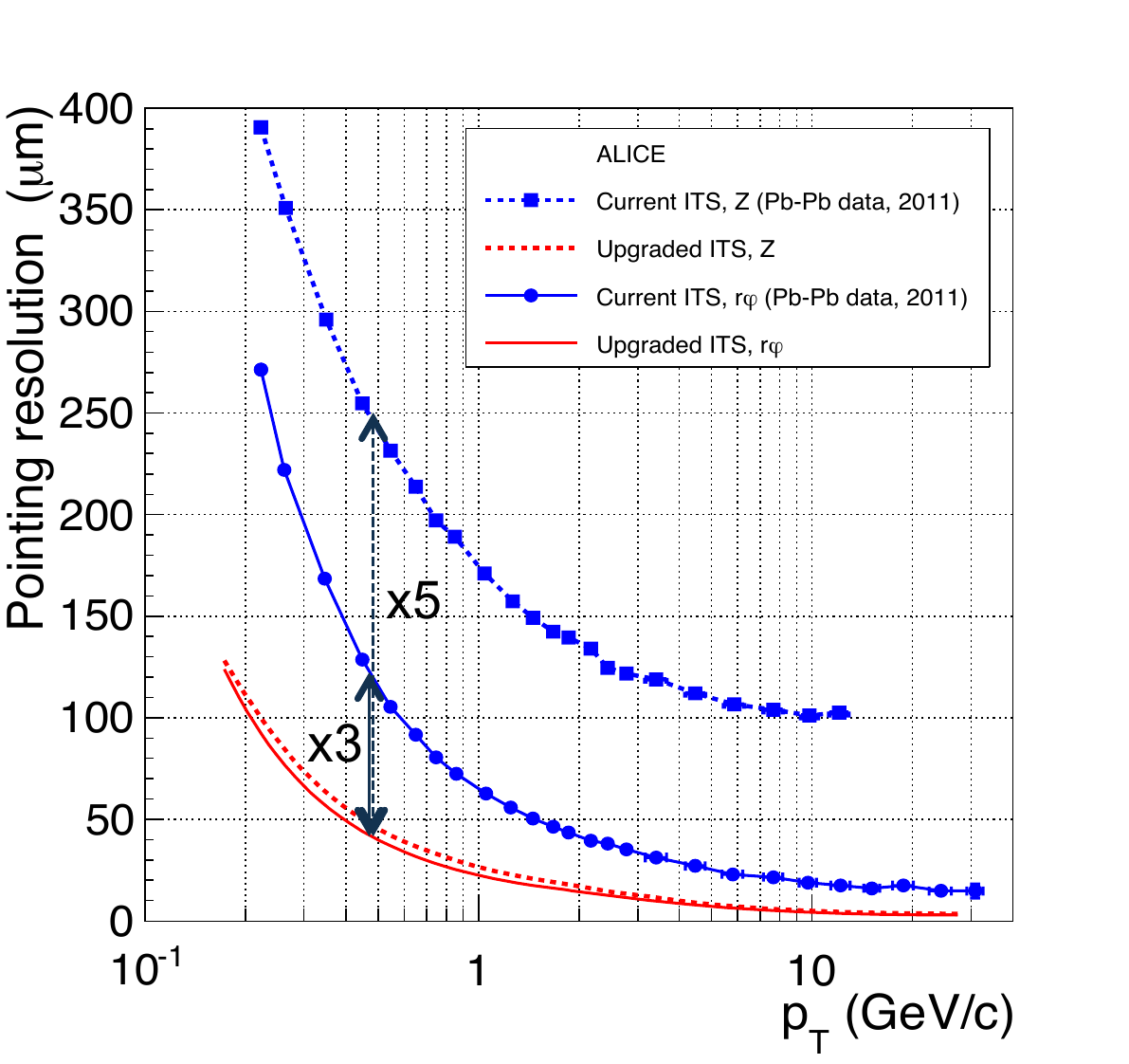}
    \put(160,815){\scriptsize $z$}
    \put(150,615){\scriptsize $r\varphi$}
  \end{overpic}
  \begin{overpic}[width=0.48\textwidth]{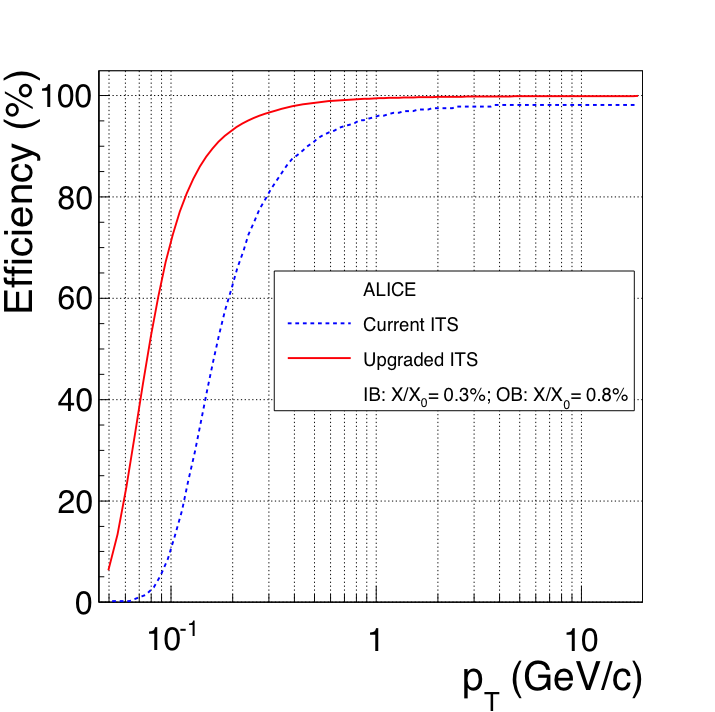}
  \end{overpic}
  \caption{Simulated pointing resolution (left) and tracking efficiency (right) of the upgraded ITS, taken from~\cite{Alice2014b}.}
  \label{fig:MC}
\end{figure}

The main goals of the ITS upgrade are to achieve an improved reconstruction of the primary vertex as well as decay vertices originating from heavy-flavour hadrons and an improved performance for the detection of low-$p_\text{T}$ particles.
The design objectives are to improve the impact parameter resolution by a factor of 3 and 5 in the $r\varphi$ and $z$ coordinate, respectively, at a $p_\text{T}$ of $\SI{500}{\MeV/\mathit{c}}$. Furthermore, the tracking efficiency and the $p_\text{T}$ resolution at low $p_\text{T}$ will improve. Corresponding Monte-Carlo simulations are shown in Fig.~\ref{fig:MC}. Additionally, the readout rate will be increased to \SI{50}{kHz} in \mbox{Pb-Pb} and \SI{400}{kHz} in pp collisions. In order to achieve this the following measures will be taken. The innermost detector layer will be moved closer to the iteraction point from $\SI{39}{\milli\metre}$ to $\SI{22}{\milli\metre}$. The material budget will be reduced down to \SI{0.3}{\%~X/X_0} per layer for the innermost layers while for the outer layers will be about \SI{0.9}{\%~X/X_0}. In addition the granularity will be increased by an additional seventh layer and by shrinking the pixel size from currently $\SI{50}{\micro\metre}\times\SI{425}{\micro\metre}$ to $\text{O}(\SI{30}{\micro\metre}\times\SI{30}{\micro\metre})$. All layers of the upgraded ITS will be equipped with pixel sensors. The upgraded ITS is designed such as to allow easy removal and insertion during the yearly shutdown periods.

\subsection{Layout and Running Environment of the Upgraded ITS}
\begin{figure}[t]
  \begin{overpic}[width=0.5\textwidth]{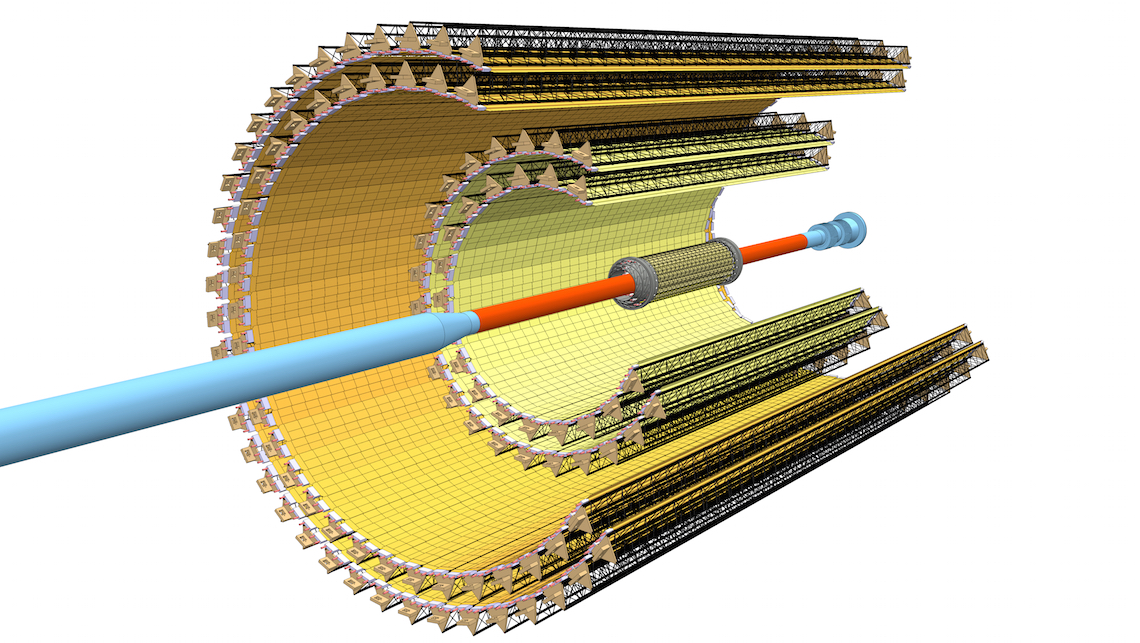}
    \put(   0,100){\footnotesize Beam pipe}
    \put( 810,490){\footnotesize Outer layers}
    \put( 750,420){\footnotesize Middle layers}
    \put( 780,320){\footnotesize Inner layers}
  \end{overpic}
  \hspace{0.1cm}
  \includegraphics[width=0.5\textwidth]{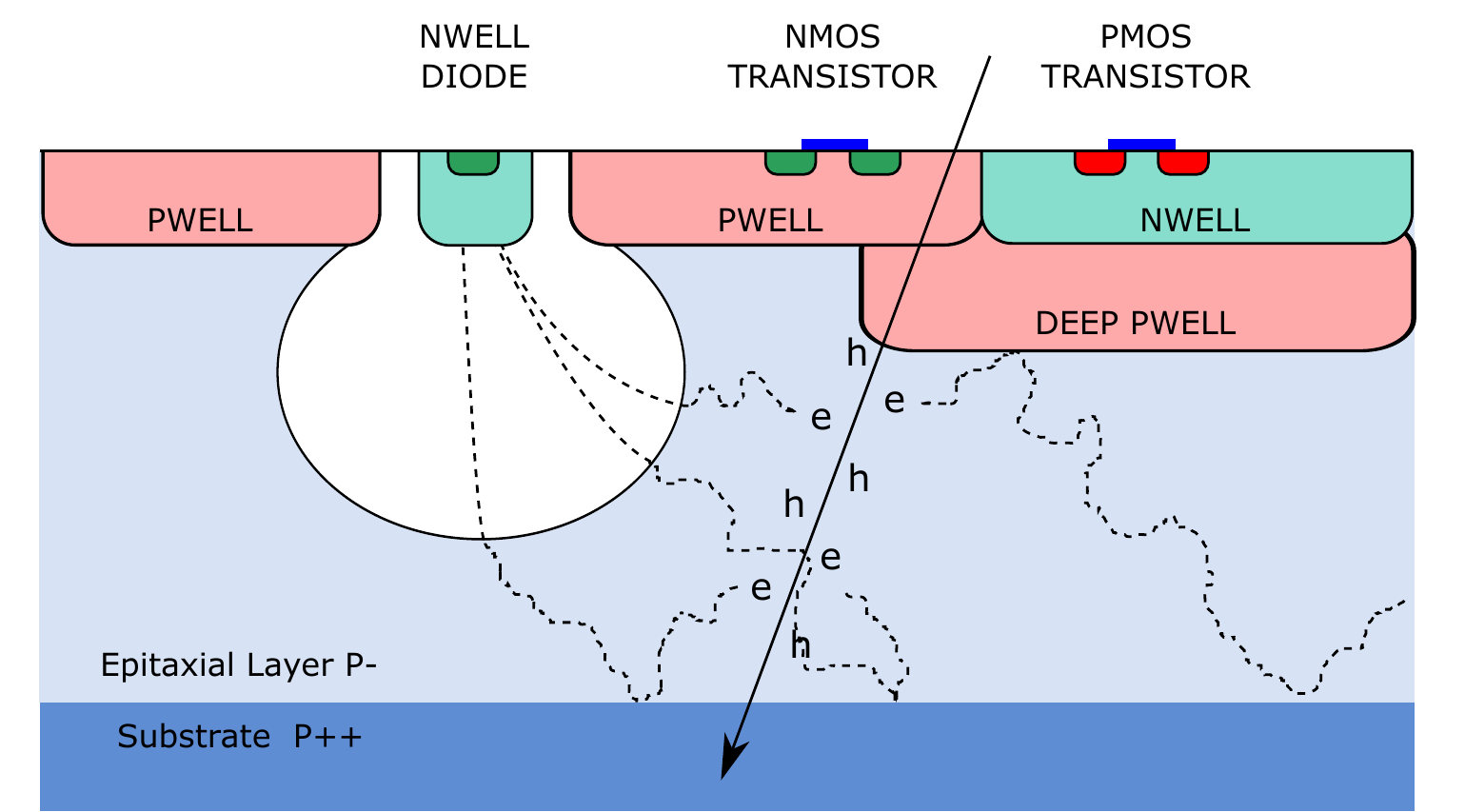}
  \caption{Layout of the upgraded ITS (left) and schematic cross section of a pixel of a monolithic silicon pixel sensor using the TowerJazz CMOS Imaging process (right), both taken from~\cite{Alice2014b}.}
  \label{fig:ITSlayout}
  \label{fig:MAPS}
\end{figure}
The upgraded ITS, as shown in Fig.~\ref{fig:ITSlayout}, will have seven layers which are grouped into the Inner Barrel, containing the innermost three layers, and the Outer Barrel containing the middle two and the outer two layers. The radii the layers are \SIlist{22;31;39}{\mm} and \SIlist{194;247;353;405}{\mm}, respectively. Although the requirements on the pixel chips are slightly different for the Inner Barrel and Outer Barrel, we aim to deploy the same chip on all seven layers (cf.\ Tab.~\ref{tab:ITSrequirements}). The upgraded ITS will provide pseudo-rapidity coverage of $|\eta|<\num{1.22}$ for $\SI{90}{\%}$ of the most luminous beam interaction region. The radial positions of the layers were optimised in order to achieve the best combined performance in terms of pointing resolution, $p_\text{T}$ resolution and tracking efficiency in Pb-Pb collisions at hit densities of about \SI{19}{\cm^{-2}\per}event on average for minimum-bias events in the innermost layer. The detector will cover a total surface of \SI{10.3}{m^{2}} containing about \num{12.5e9} pixels with binary readout. The upgraded ITS will be operated at room temperature (\SIrange{20}{30}{\degreeCelsius}) using water cooling. The expected radiation load at the innermost layer is expected to be \SI{700}{krad} of Total Ionising Dose (TID) and \SI{1e13}{\SI{1}{MeV~n_{eq}/\cm\squared}} of Non-Ionising Energy Loss (NIEL) including a safety factor of ten. In order to meet the material budget requirements the silicon sensors will be thinned down to \SI{50}{\um}.
\begin{table}[t]
  \caption{General pixel-chip requirements~\cite{Alice2014b}.}\label{tab:ITSrequirements}
  \centering
  \begin{tabular}{lcc}
    \toprule
    \textbf{Parameter}           & \textbf{Inner Barrel} & \textbf{Outer Barrel} \\ \midrule
    Chip dimensions              & \multicolumn{2}{c}{$\SI{15}{\mm}\times\SI{30}{\mm}$ ($r\varphi\times z$)} \\
    Sensor thickness             & \multicolumn{2}{c}{\SI{50}{\um}} \\
    Spatial resolution           & \SI{5}{\um}           & \SI{10}{\um} \\
    Detection efficiency         & \multicolumn{2}{c}{$>$\,\SI{99}{\%}} \\
    Fake hit rate                & \multicolumn{2}{c}{$<10^{-5}$\,$\text{event}^{-1}\text{pixel}^{-1}$} \\
    Integration time             & \multicolumn{2}{c}{$<$\,\SI{30}{\us}} \\
    Power density                & $<\SI{300}{\mW/\cm\squared}$ & $<\SI{100}{\mW/\cm\squared}$ \\
    Temperature                  & \multicolumn{2}{c}{\SIrange{20}{30}{\degreeCelsius}} \\
    TID radiation hardness$^{a}$ & \SI{700}{krad} & \SI{10}{krad} \\
    NIEL radiation hardness$^{a}$ & \SI{1e13}{\SI{1}{MeV~n_{eq}/\cm\squared}} & \SI{3e10}{\SI{1}{MeV~n_{eq}/\cm\squared}} \\ \bottomrule
  \end{tabular} \\
  { \footnotesize $^{a}$These values include a safety factor of ten.}
\end{table}
\subsection{Choice of Pixel Chip Technology}
Summarising the considerations above, the upgraded ITS will have very thin sensors, very high granularity and will cover a fairly large area. Furthermore, the radiation levels are only moderate compared to the other LHC experiments. In the past decade there has been a lot of progress on Monolithic Active Pixels Sensors (MAPS), which can now be considered for the construction of tracking systems in high-energy physics experiments. MAPS allow for very thin sensors, as a single die is used as detection volume and for the readout electronics. Additionally, no bump bonding or similar interconnection of detection and readout chip are needed, and this interconnection usually limits cost and pixel density. The ULTIMATE chip of the STAR PXL detector~\cite{Greiner2011} at RHIC is the first successfully running, large-scale application. However, further R\&D is required to meet the much more stringent requirements of the ITS upgrade compared to the STAR experiment in terms of integration time, power consumption and radiation hardness.

\subsection{Pixel Chip Development}
The sensors of  upgraded ITS will be manufactured using the \SI{0.18}{\um} CMOS Imaging Sensor process by TowerJazz~\cite{TowerJazz}. This process provides up to six metal layers allowing for a high-density, low-power circuitry. Furthermore, the gate oxide thickness of about \SI{3}{\nm} provides a sufficient TID radiation tolerance. This has already been confirmed in measurements on basic transistor structures~\cite{Hillemanns2013}. The key feature of the process, however, is the special deep p-well. As shown in Fig.~\ref{fig:MAPS}, the n-wells of PMOS transistors are housed in additional p-wells, preventing the transistor \mbox{n-wells} from competing with the n-well of the collection electrode for charge collection. Hence, full CMOS logic can be used within the matrix and as consequence, more complex in-pixel circuitry is possible. An epitaxial layer with high resistivity $(\sim\si{k\ohm\cm})$ serves as active volume. In order to increase the depletion volume and to optimise the charge-collection efficiency, a moderate reverse substrate bias can be applied. This is essential to increase the output signal of the collection n-well which is proportional to $\sim{Q/C}$. In order to achieve a high signal, the charge collected by the central pixel needs to be increased. Furthermore, the capacitance of the pixel needs to be minimised by shrinking the diode surface and increasing the depletion volume which is supported by additional reverse substrate bias. Achieving a good $Q/C$-ratio leads to an improved signal-to-noise ratio and as a consequence also to a less power consuming design of the circuitry.

\subsection{Pixel Chip Architecture}
\begin{figure}
\includegraphics[width=.5\textwidth]{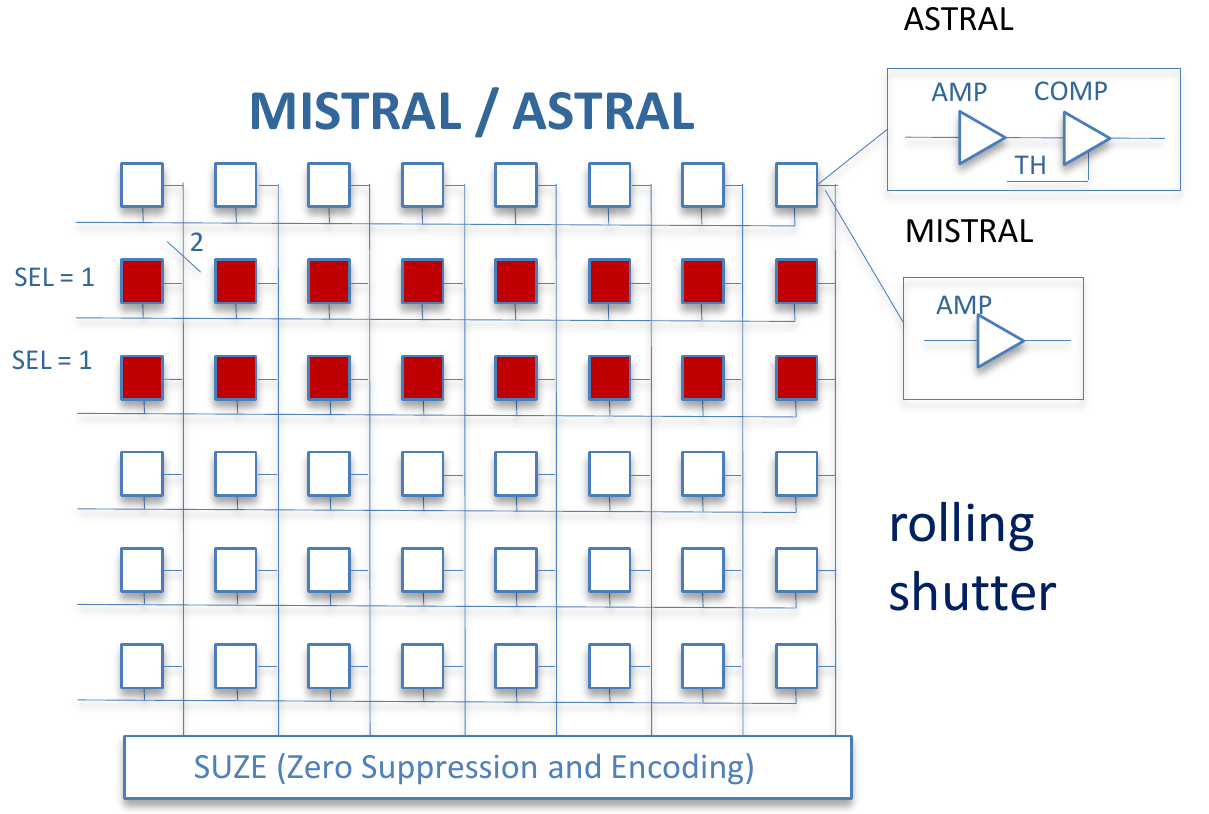}
\includegraphics[width=.5\textwidth]{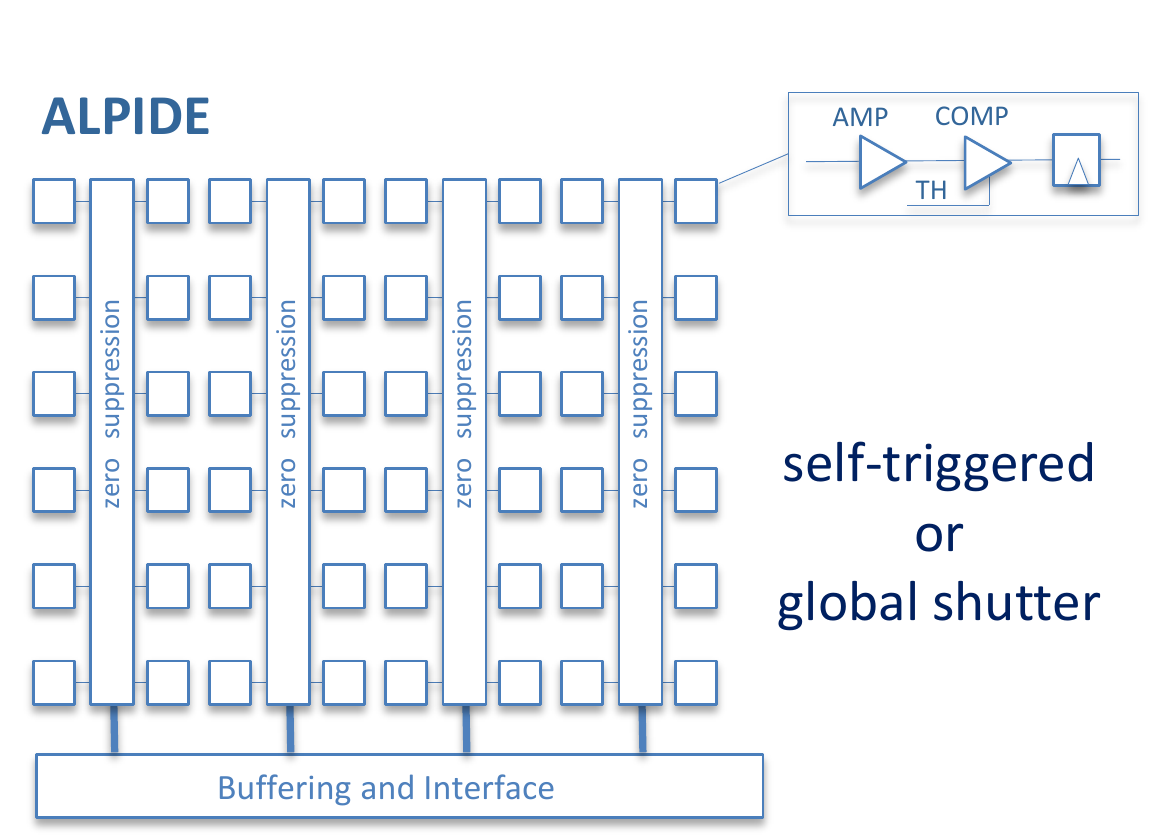}
\caption{Sketch of the architectures deployed in the ASTRAL (left) and MISTRAL (right) design stream.}\label{fig:architectures}
\end{figure}
Currently, two R\&D design streams are under development, called ASTRAL and ALPIDE (cf.~Fig.~\ref{fig:architectures}).
The ASTRAL chip is based on a rolling-shutter architecture, where rows of pixels are read simultaneously and the integration time is defined by the time the shutter needs to return to the same row. ASTRAL contains in-pixel discriminators and end-of-column sparsification called SUZE. This architecture is based on the ULTIMATE chip developed for the STAR PXL detector. The MISTRAL is a variant of ASTRAL having end-of-column discriminators, resulting in a simpler in-pixel circuitry but a higher power density.

On the other hand ALPIDE operates in a global shutter mode. The shutter can either be started by an external trigger signal or be kept continuously open. In continuous-integration mode the shutter is only closed to advance to the next event. The chip features in-pixel discrimination and in-pixel hit buffers. These buffers allow to acquire consecutive events while the readout of the previous event is still ongoing. Furthermore, a priority encoder is used to achieve in-matrix sparsification. Only pixels containing hits are propagated to the end-of-column further reducing the power-consumption and the area necessary for the peripheral logic.

After the successful development and characterisation of small-scale prototypes, large-scale prototypes, close to the final chip dimensions, of both ASTRAL and ALPIDE architectures are currently being characterised.
\subsection{ASTRAL/MISTRAL Prototypes}
\begin{figure}
  \centering
  \includegraphics[width=0.24\textwidth]{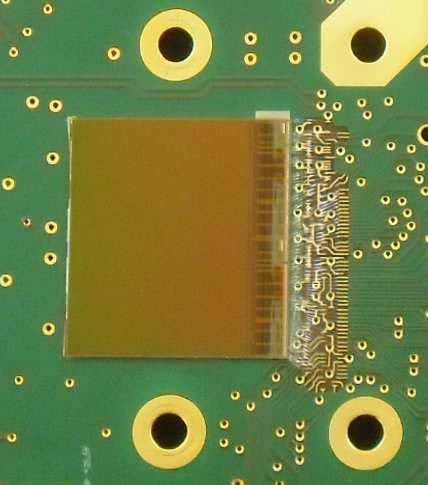}
  \begin{overpic}[width=0.5\textwidth]{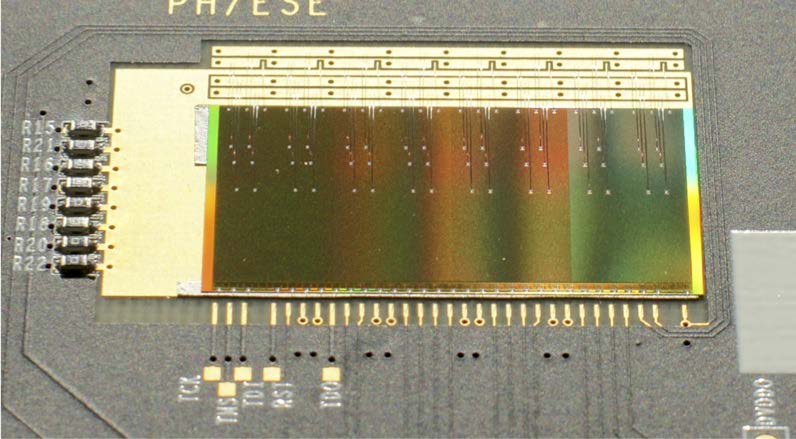}
  \end{overpic}
  \caption{Two FSBB-M0 chips arranged in a column (left) and pALPIDEfs chip (right).}
  \label{fig:Chips}
\end{figure}
The FSBB-M0 (cf.~Fig.~\ref{fig:Chips}) is a Full-Scale Building Block of the MISTRAL design stream. Three \mbox{FSBBs} form a full chip. The FSBB-M0 chip features $416 \times 416$ pixels of \mbox{$\SI{22}{\um}\times\SI{33}{\um}$} which are read out by double-row end-of-column discriminators. The integration time of this prototype is \SI{40}{\us}. The FSBB-A0 is an implementation of the ASTRAL version of the FSBB achieving \SI{20}{\us} integration time deploying in-pixel discrimination.
\begin{figure}[b]
  \begin{overpic}[width=0.48\textwidth]{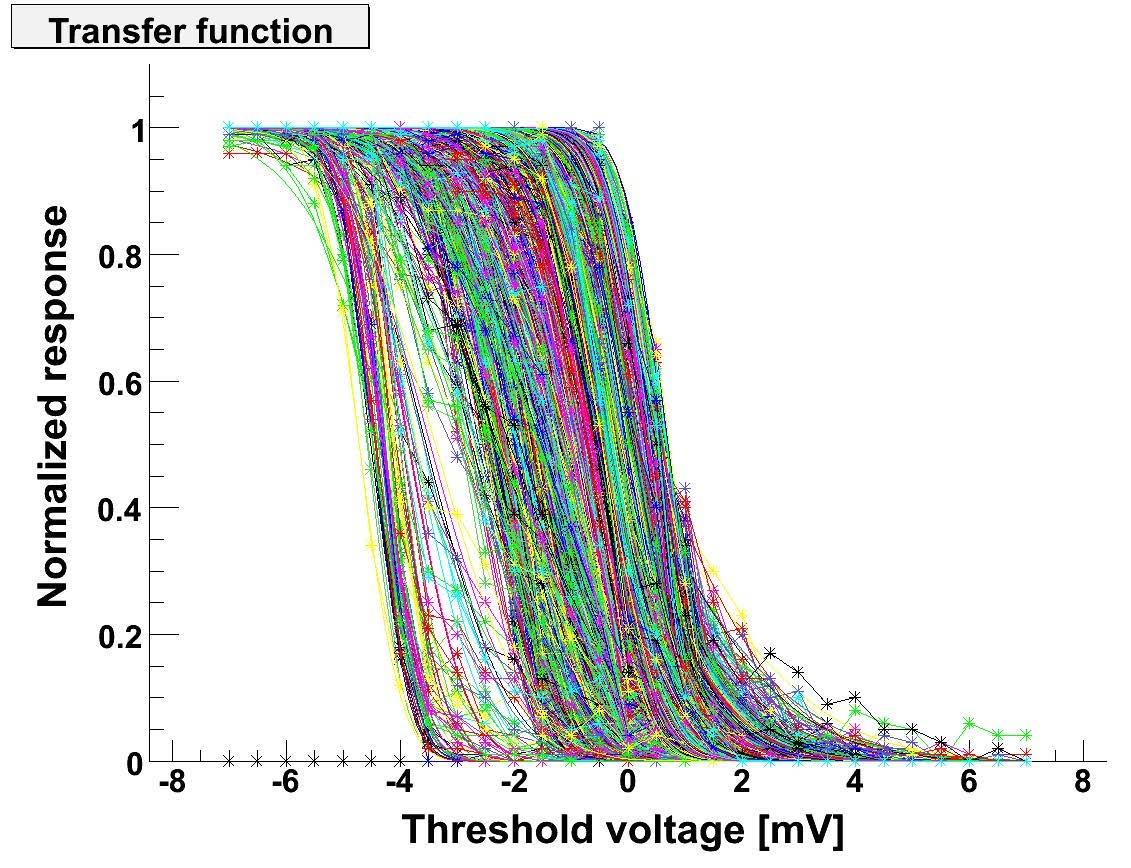}
  \end{overpic}
  \begin{overpic}[width=0.50\textwidth]{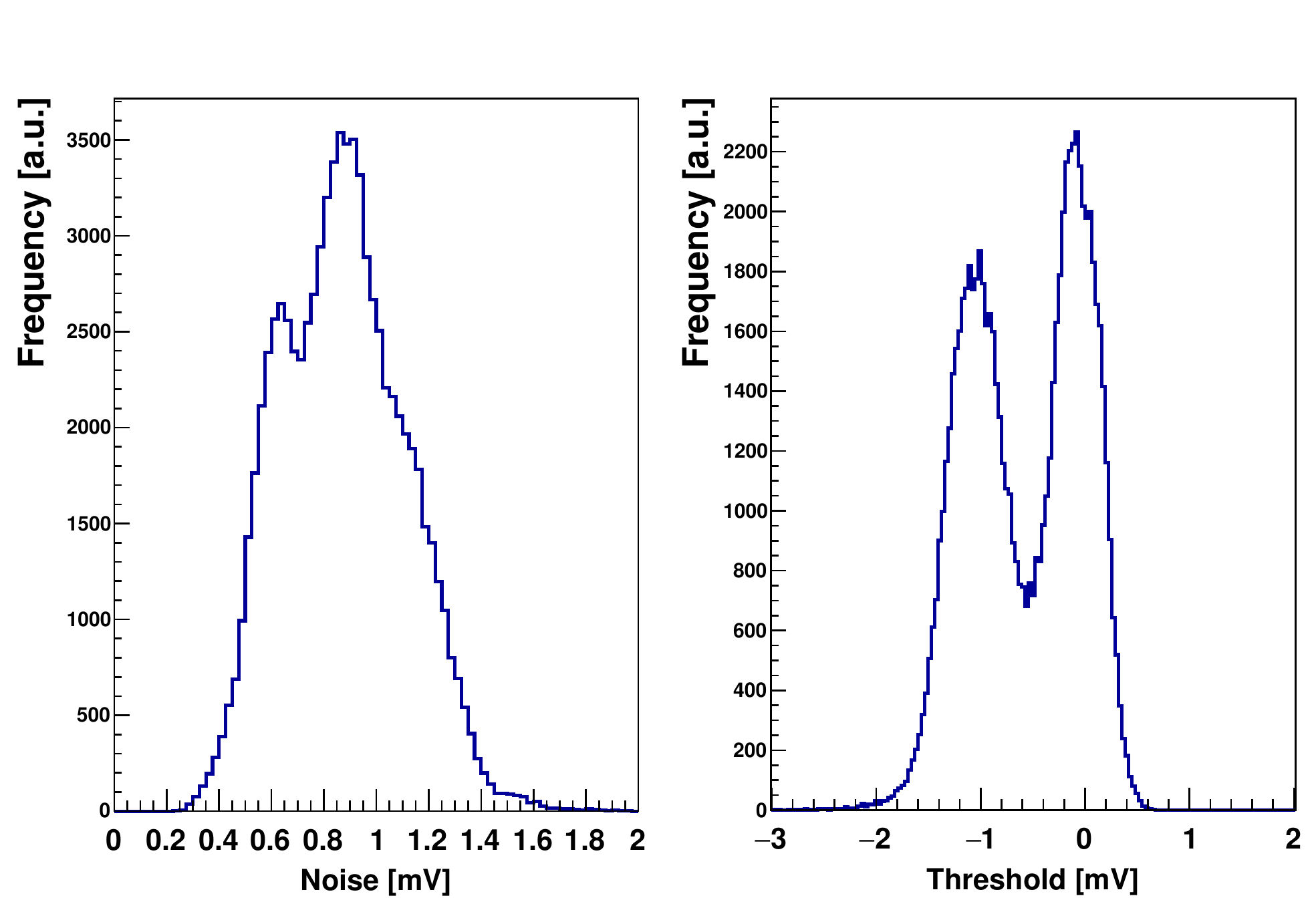}
    \put( 30,700){\footnotesize $\text{Temporal Noise}\approx\SI{0.87}{\mV}$}
    \put(530,700){\footnotesize $\text{Fixed-Pattern Noise}\approx\SI{0.55}{\mV}$}
  \end{overpic}
  \caption{Discriminator transfer function (left), Temporal Noise distribution (centre) and Fixed Pattern Noise (right) of an example FSBB-M0.}
  \label{fig:FSBB-M0}
\end{figure}
A total of 25 FSBB-M0 have been extensively characterised in the laboratory, showing similar noise performance. The corresponding transfer function for the variation of the discriminator threshold is shown in Fig.~\ref{fig:FSBB-M0}. A Temporal Noise (TN), the time-like threshold dispersion, of \SI{0.87}{\mV} and the Fixed Pattern Noise (FPN), the spatial spread of the pixel threshold, of \SI{0.55}{\mV} were achieved. The double-peak structures are due to cross-coupling. Further tests at the CERN SPS are planned for the near future.
\subsection{pALPIDEfs - a Full-Scale Prototype of the ALPIDE}
The pALPIDEfs is a full-scale prototype of the ALPIDE family (cf.~Fig.~\ref{fig:Chips}) with a dimension of $\SI{30}{\mm}\times\SI{15}{\mm}$ containing about \num{5e5} pixels of $\SI{28}{\um}\times\SI{28}{\um}$. The power consumption per pixel front-end is \SI{40}{nW} leading to a power density of $\SI{4.7}{\mW/\cm\squared}$. In order to increase the depletion volume, a reverse substrate bias can be applied to this chip. In its current version, four sectors containing different design options are implemented and in particular, several diode geometries with different sizes of p-well openings and reset mechanisms, namely PMOS resets and diode resets as outlined in Tab.~\ref{tab:pALPIDEfsPixels}. The pALPIDEfs features in-matrix sparsification based on a priority encoder. The target power density for future prototypes of this family excluding off-chip data transmission is about \SI{30}{\mW/\cm\squared}.
\begin{table}[t]
  \centering
  \caption{pALPIDEfs pixel properties.}\label{tab:pALPIDEfsPixels}
  \begin{tabular}{ccccc}
    \toprule
    Sector & n-well      & Spacing     & p-well       & Reset \\
           & diameter    &             & opening      &       \\ \midrule
    0      & \SI{2}{\um} & \SI{1}{\um} & \SI{4}{\um}  & PMOS  \\
    1      & \SI{2}{\um} & \SI{2}{\um} & \SI{6}{\um}  & PMOS  \\
    2      & \SI{2}{\um} & \SI{2}{\um} & \SI{6}{\um}  & Diode \\
    3      & \SI{2}{\um} & \SI{4}{\um} & \SI{10}{\um} & PMOS  \\ \bottomrule
  \end{tabular}
  \vspace{0.3cm}
\end{table}
\begin{figure}[b]
  \centering
  \includegraphics[width=0.48\textwidth]{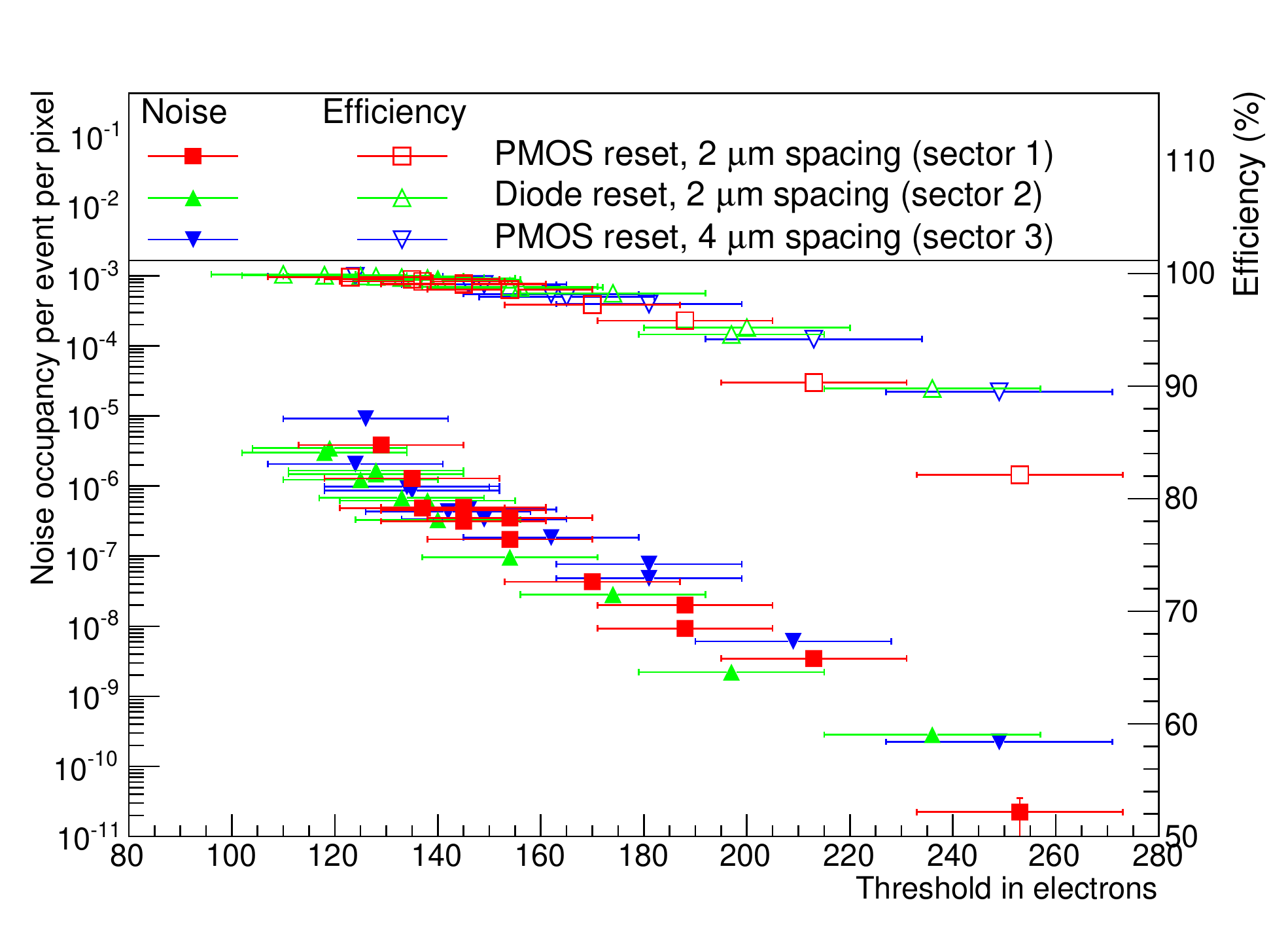}
  \includegraphics[width=0.5\textwidth]{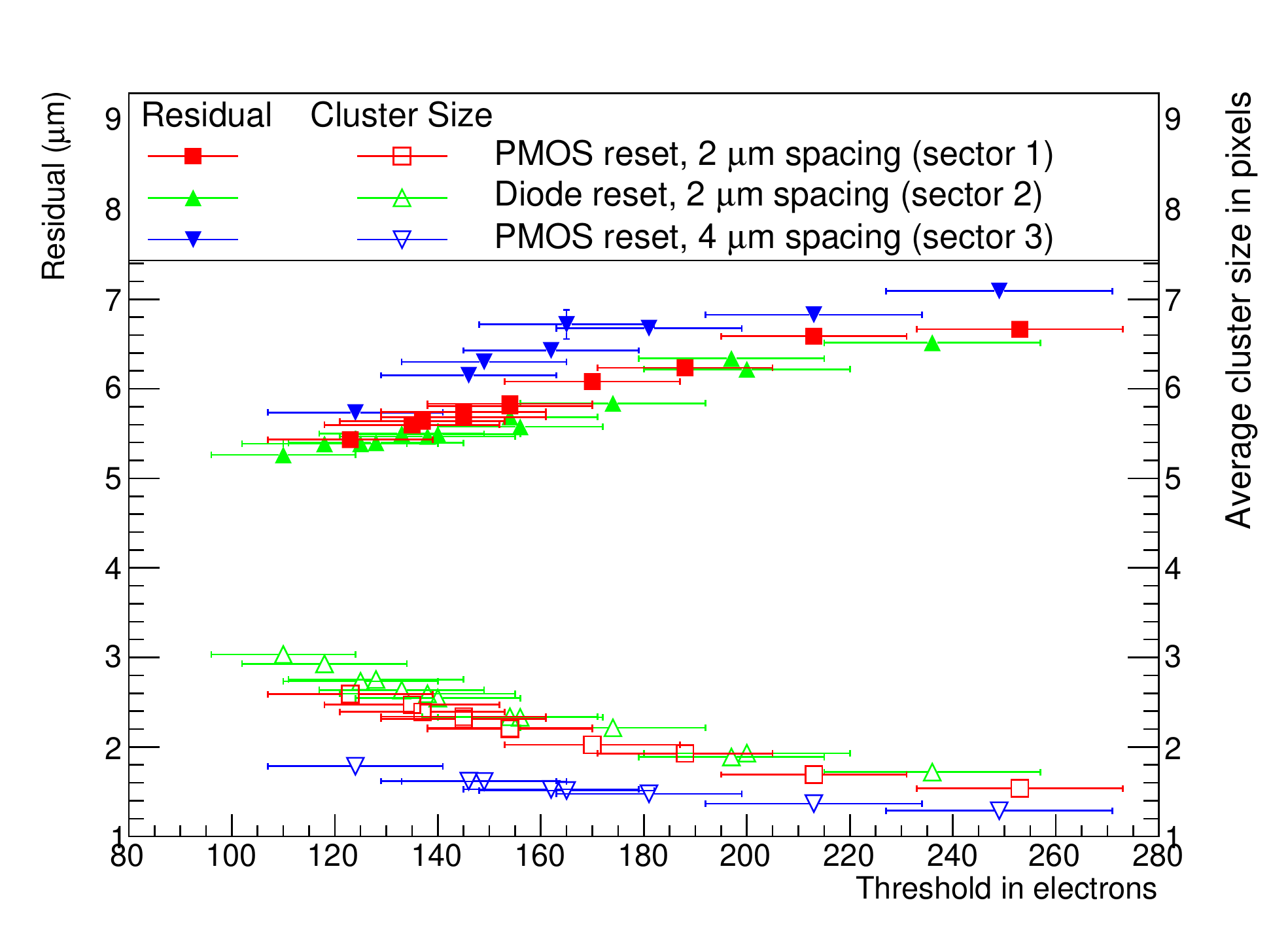}
  \caption{Detection efficiency (open symbols) and noise (full symbols) of the pALPIDEfs (left) and uncorrected position resolution (full symbols) as well as cluster size (open symbols) of the pALPIDEfs (right), both without the application of reverse substrate bias. The sectors 1, 2, and 3 are drawn with red, green and blue symbols, respectively.}
  \label{fig:detEffNoiseResClus}
\end{figure}

\begin{figure}[t]
  \centering
  \includegraphics[width=0.48\textwidth]{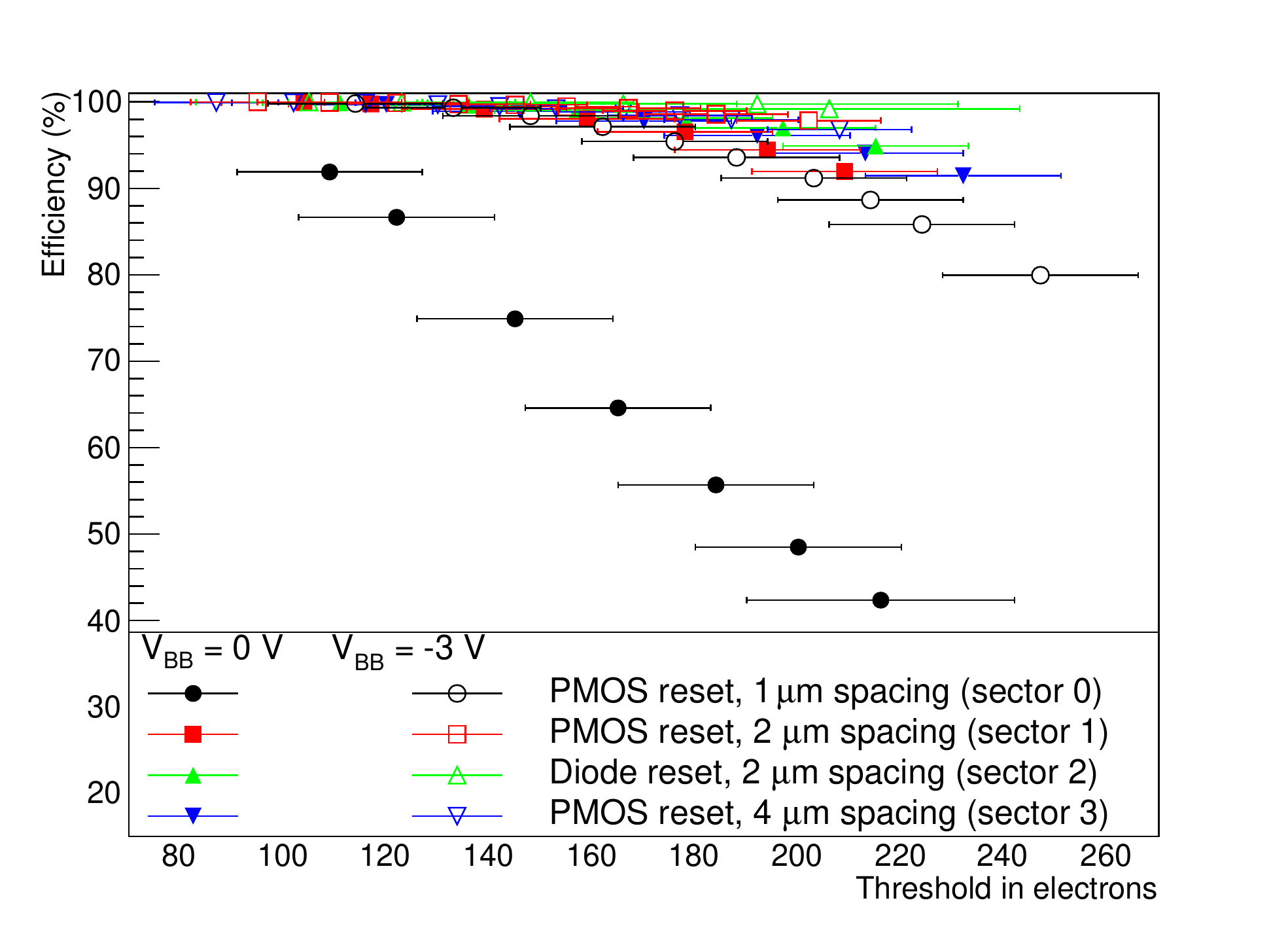}
  \includegraphics[width=0.48\textwidth]{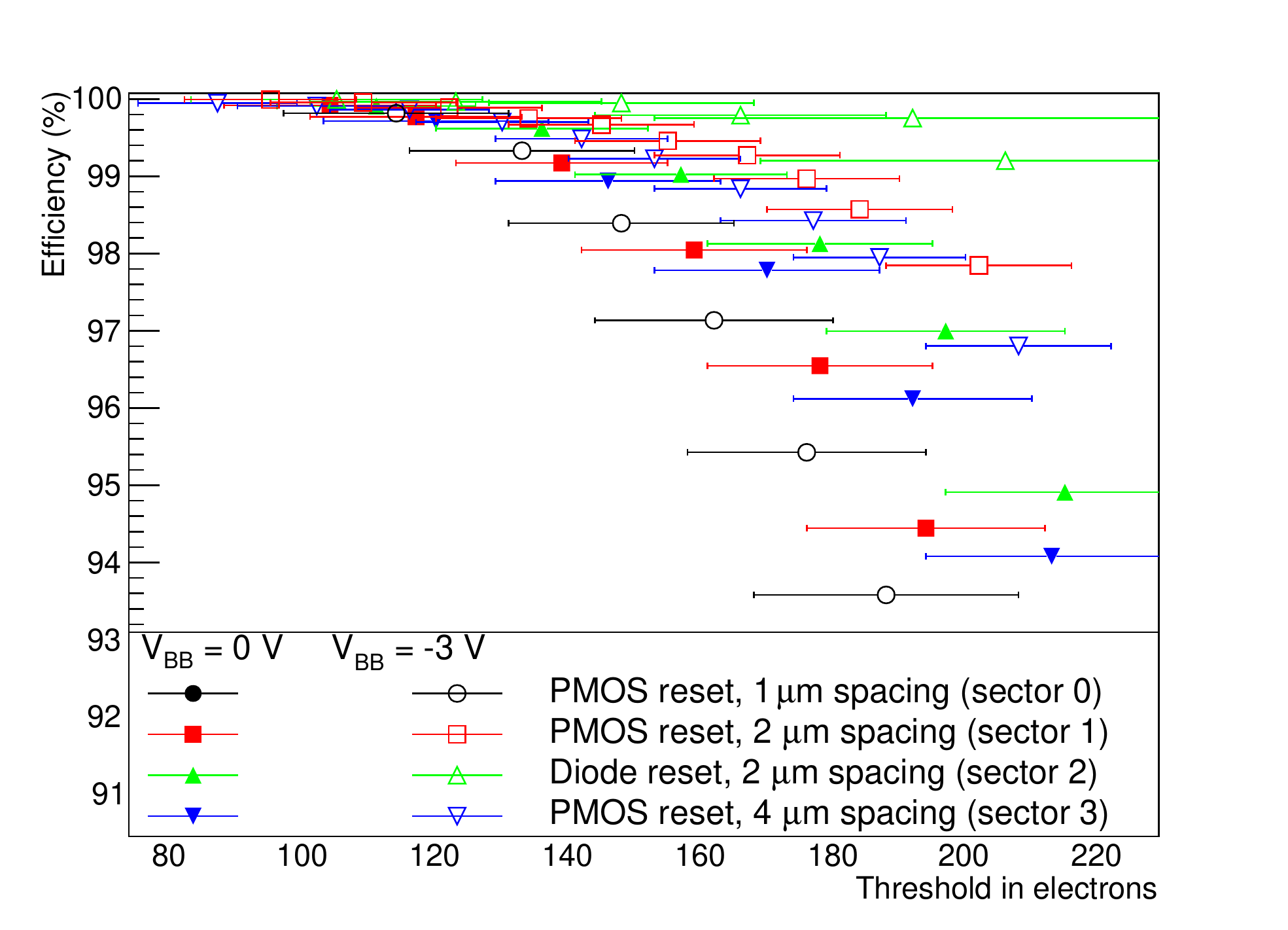}
  \caption{Influence of reverse substrate bias on the detection efficiency of the pALPIDEfs (left) and zoomed view (right). Open symbols are with a reverse substrate bias of $V_\text{BB}=\SI{-3}{V}$ and full symbols without reverse substrate bias. The sectors 0, 1, 2, and 3 are drawn with black, red, green and blue symbols, respectively.}
  \label{fig:BB}
\end{figure}

The pALPIDEfs has been characterised in the laboratory as well as in test beam. The following results were obtained using a telescope of 6 or 7 pALPIDEfs at the CERN PS using a \SI{6}{GeV} pion beam. The first results on detection efficiency and noise are shown in Fig.~\ref{fig:detEffNoiseResClus}~(left). A detection efficiency of \SI{99}{\%} at a fake-hit rate of $10^{-5}$ was measured. The results were obtained having only 20 pixels masked.
The position resolution has been measured and the corresponding residuals and the cluster size are shown in Fig.~\ref{fig:detEffNoiseResClus}~(right). This measurement still includes a tracking error of about \SI{3}{\um} leading nevertheless to an uncorrected spatial resolution of about \SI{5.5}{\um}.
In Fig.~\ref{fig:detEffNoiseResClus}, sector 0 has been excluded. This sector reaches a detection efficiency of above \SI{99}{\%} only using reverse substrate bias as shown in Fig.~\ref{fig:BB}. Also the other sectors gain margin using reverse substrate bias, allowing for higher thresholds maintaining detection efficiencies above \SI{99}{\%}.
All results confirm the positive effect of a larger spacing on the detection performance. This can most likely be attributed to an increase depleted volume due to decreased side-wall capacitance. Furthermore, pixels with diode reset as in sector 2 perform better than the pixels with PMOS reset.
\section{Mechanics and Assembly}
\begin{figure}
  \centering
  \includegraphics[width=.39\textwidth]{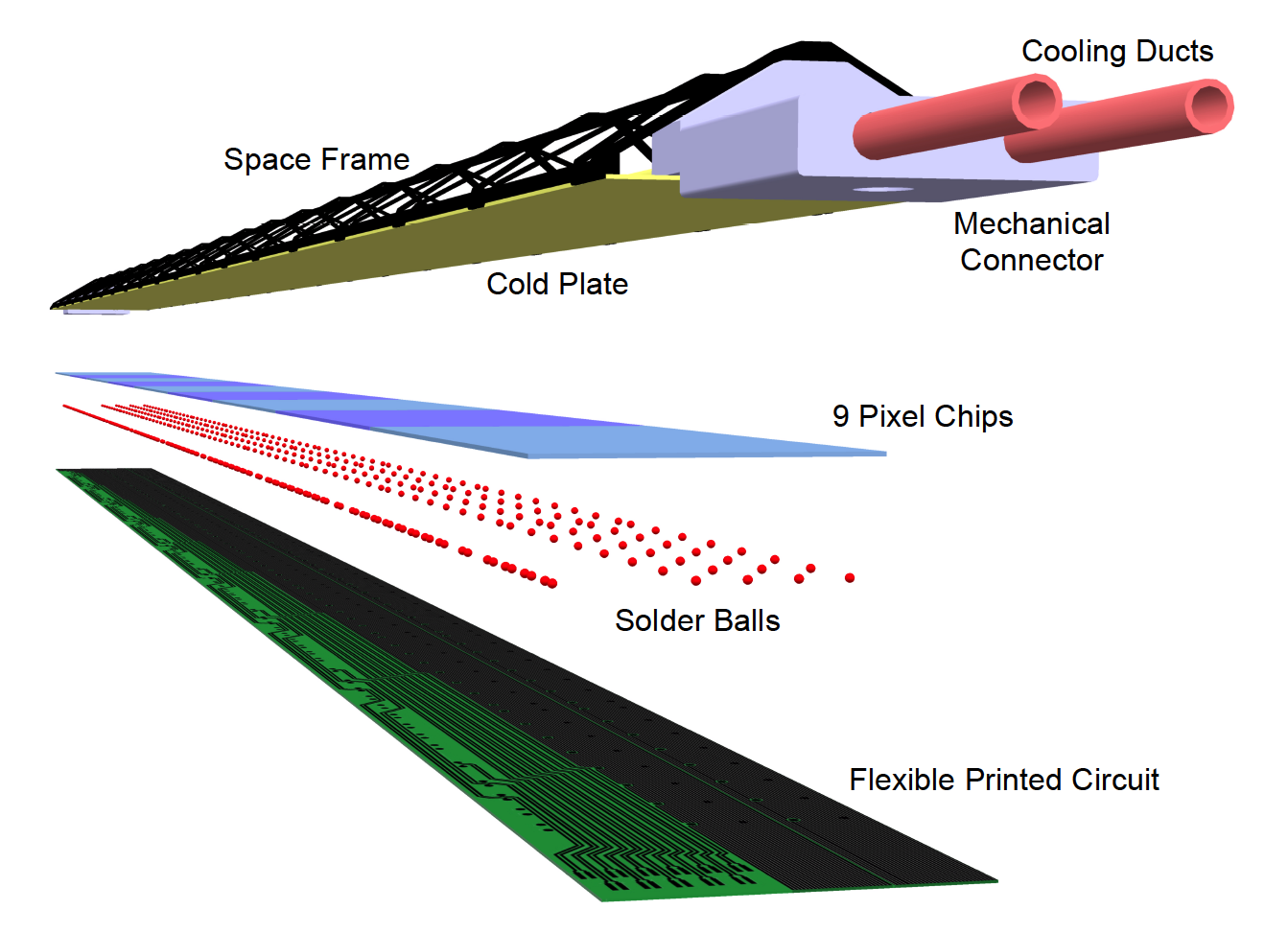}
  \includegraphics[width=.59\textwidth]{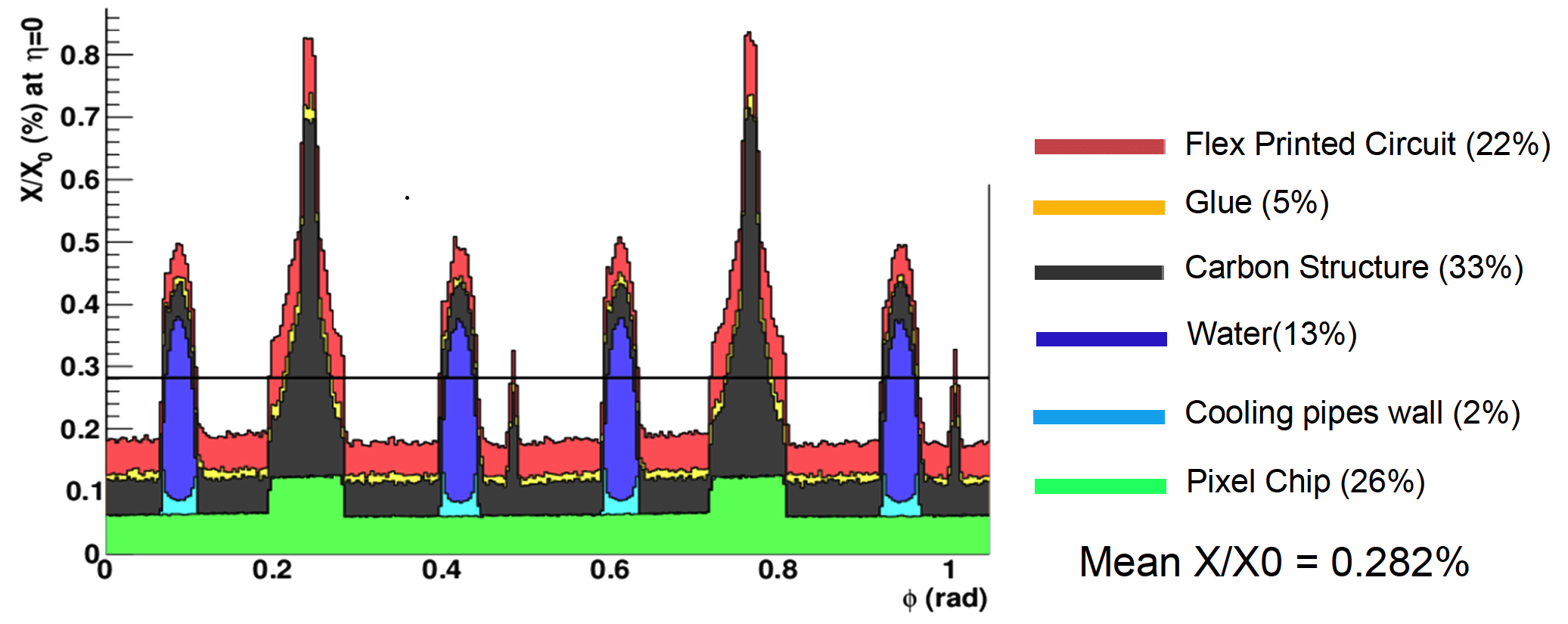}
  \caption{Material budget distribution of the inner barrel (left), exploded schematic view of an inner barrel stave (right).}
  \label{fig:Mechanics}
\end{figure}

As mentioned above, the upgraded ITS consists of an Inner Barrel (IB) and an \mbox{Outer Barrel (OB)}. The basic element of a layer is the stave, which consists of a carbon space frame to which the cold plate and the cooling ducts are attached. Above the cold plate a number of pixel chips, 9 for the IB and 14 for the OB, connected to a common Flexible Printed Circuit (FPC) are glued (cf.~Fig.~\ref{fig:Mechanics}, left). The FPC consists of a polyimide with a low thermal expansion coefficient plus aluminium and copper as conductor for the IB and OB, respectively. The chip will be connected to the FPC using laser soldering, allowing a distribution of the connection pads over the entire chip surface rather than its periphery. This has been successfully prototyped and is working on pALPIDEfs chips. The staves of the IB will have a length of \SI{270}{\mm}. The staves of the middle and outer two layers of the OB will be \SI{843}{\mm} and \SI{1475}{\mm} long. In order to improve the pointing resolution, the material budget of the inner layers will not exceed \SI{0.3}{\%~X/X_{0}} on average per layer. The material budget will be higher in the regions of stave overlap and cooling pipes (cf.~Fig.~\ref{fig:Mechanics}, right).

For the OB and its wider and longer staves further segmentations are introduced. An OB stave consists of two half-staves. The half-staves of the same stave as well as the half-staves of adjacent staves are overlapping in order to minimise the dead area. The material budget is \SI{0.9}{\%~X/X_{0}} per layer.

First prototypes of the IB and OB mechanics have been assembled and have been successfully characterised for their mechanical strength and thermal properties~\cite{Alice2014b}.

\section{Summary and Outlook}
ALICE will replace the entire ITS by a MAPS-based, pixel-only tracker in 2019. This upgrade will significantly improve impact-parameter resolution and increase readout-rate capabilities. Moreover, better tracking efficiency and $p_\text{T}$ resolution at low $p_\text{T}$ will be achieved. For the pixel sensor R\&D, large-scale prototypes of two separate design architectures are currently being characterised and have shown satisfactory results. The mechanical structures have been prototyped and successfully tested. Also the novel laser soldering to establish the connection between the pixel chips and the flexible printed circuit has been successfully applied to working prototype chips. Further close-to-final design prototypes will be tested in the near future and assembled in staves in order to prepare the mass production.


\begin{thebibliography}{99}
\bibitem{Aamodt2008}
K. Aamodt et al. \emph{The ALICE experiment at the CERN LHC}. In: \emph{JINST} 3 (2008) S08002.
\bibitem{Alice2014a}
The ALICE Collaboration. \emph{Upgrade of the ALICE Experiment: Letter Of Intent}. In \emph{J.Phys. G}41 (2014) 087001 [CERN-LHCC-2012-012. CERN-LHCC-I-022. ALICE-UG-001].
\bibitem{Alice2012}
The ALICE Collaboration. \emph{Conceptual Design Report for the Upgrade of the ALICE ITS}. Tech. rep. CERN-LHCC-2012-005. LHCC-G-159. Geneva: CERN, Mar. 2012.
\bibitem{Alice2014b}
The ALICE Collaboration. \emph{Technical Design Report for the Upgrade of the ALICE Inner Tracking System}. In \emph{J.Phys. G}41 (2014) 087002 [CERN-LHCC-2013-024. ALICE-TDR-017].
\bibitem{Alice2013a}
The ALICE Collaboration. \emph{Addendum of the Letter Of Intent for the Upgrade of the ALICE Experiment : The Muon Forward Tracker}. Tech. rep. CERN-LHCC-2013-014. LHCC-I-022- ADD-1. Geneva: CERN, Aug. 2013.
\bibitem{Alice2013b}
The ALICE Collaboration. \emph{Upgrade of the ALICE Time Projection Chamber}. Tech. rep. CERN-LHCC-2013-020. ALICE-TDR-016. Geneva: CERN, Oct. 2013.
\bibitem{Alice2013c}
The ALICE Collaboration. \emph{Upgrade of the ALICE Readout \& Trigger System}. Tech. rep. CERN-LHCC-2013-019. ALICE-TDR-015. Geneva: CERN, Sept. 2013.
%\bibitem{Alice2014c}
%The ALICE Collaboration. \emph{Technical Design Report for the Upgrade of the Online–Offline System%}. Tech. rep. under preparation
\bibitem{Greiner2011}
L. Greiner et al. \emph{A MAPS based vertex detector for the STAR experiment at RHIC}. In: \emph{NIM A} 650.1 (2011). International Workshop on Semiconductor Pixel Detectors for Particles and Imaging 2010, pp. 68 –72.
\bibitem{TowerJazz}
Tower Jazz. \emph{www.jazzsemi.com}.
\bibitem{Hillemanns2013}
H. Hillemanns et al. \emph{Radiation hardness and detector performance of new 180nm CMOS MAPS prototype test structures developed for the upgrade of the ALICE Inner Tracking System}. In Proceedings NSSMIC C13-10-26 (2013).
\end{thebibliography}
\end{document}